\documentclass[reprint,aps,prl,twocolumn,showpacs,footnoteinbib]{revtex4-1}
\usepackage{amssymb}
\usepackage{amsmath}
\usepackage{graphicx}
\begin{document}
\title{Signatures of Bloch-band geometry on excitons:  non-hydrogenic spectra in transition metal dichalcogenides}
\author{Ajit Srivastava}
\author{Atac Imamo\u{g}lu}

\affiliation{Institute of Quantum Electronics, ETH Z\"{u}rich, CH-8093
Z\"{u}rich, Switzerland.}

\date{\today }

\begin{abstract}
The geometry of electronic bands in a solid can drastically alter
single-particle charge and spin transport. We show here that
collective optical excitations arising from Coulomb interactions
also exhibit unique signatures of Berry curvature and quantum
geometric tensor. A non-zero Berry curvature mixes
and lifts the degeneracy of $l \neq 0$ states, leading to a
time-reversal-symmetric analog of the orbital Zeeman effect. The
quantum geometric tensor, on the other hand, leads to $l$-dependent
shifts of exciton states that is analogous to the Lamb shift. Our
results provide an explanation of the non-hydrogenic exciton
spectrum recently calculated for transition metal dichalcogenides.
Numerically, we find a Berry curvature induced splitting of
$\sim 10$ meV between the $2p_x \pm i2p_y$ states of WSe$_2$.
\end{abstract}

%\pacs{03.67.Lx, 73.21.La, 42.50.-p}
\maketitle

\paragraph{Introduction \textemdash}
An exciton, comprising of a bound
electron-hole pair, is an elementary optical excitation of a
semiconductor. In most semiconductors, large dielectric constant and
small effective masses of charge carriers result in a weakly bound
exciton with a Bohr-radius much larger than the lattice constant.
Such excitons are termed as Wannier-Mott excitons and play a central
role in understanding the optical response of a number of
condensed-matter
systems~\cite{Byrnes14,Smolka14}. Although the  charge carriers in a semiconductor are described by
Bloch-waves, excitonic spectra of most semiconductors resemble the
hydrogenic series consisting of Rydberg series~\cite{Kazimierczuk14}. At
a first glance, this is surprising since the wave functions of
electrons and holes in a  solid obey crystal symmetry. For example,
discrete translational symmetry in a crystal implies that
single-particle band dispersions of electrons and holes are periodic in
crystal momentum $\mathbf{k}$, necessarily making their kinetic
energy non-parabolic. Nevertheless, in a semiconductor with a direct
band gap at the $\Gamma$-point ($\mathbf{k}$ = 0) and large Bohr
radius such that exciton is made from $\mathbf{k}$-states close to
band-minima, full symmetry of vacuum is restored and a hydrogenic
spectrum for excitons is obtained.

%In addition to non-parabolic kinetic energy of single-particle states,  the dependence of Coulomb interaction between electron and hole on Bloch states can also modify the hydrogenic series. More precisely, the latter depends on how the Bloch states are arranged in Brillouin zone or on the geometry of Bloch bands. While the role of Bloch-band geometry in determining transport properties of non-interacting Bloch electrons is well established~\cite{Xiao10, Neupert13}, its role in modifying Coulomb attraction leading to bound-state formation has not been explicitly analyzed.

% I commented the above paragraph since I thought it to be a bit destruptive to the flow. Instead I included the last sentence at the end of the following paragraph

It was recently reported that in monolayers of semiconducting
transition metal dichalcogenides (TMDs) such as MoS$_2$ or WSe$_2$,
excitonic spectra deviates strongly from that of a hydrogen atom. In
particular, states with identical principal quantum number, such as
2$s$ and 2$p$ are not degenerate~\cite{Ye14, He14, Chernikov14,
Wang15, Ugeda14, Zhu15}. Moreover, the degeneracy of 2$p$ states was
also found to be lifted in tight-binding calculations~\cite{Wu15}.
These exciting predictions were attributed to non-local dielectric
screening in TMDs where exciton Bohr radius of the 1s exciton is
comparable to the monolayer thickness~\cite{Berkelbach13}. In fact,
as a consequence of the tightly bound nature of Wannier-Mott
excitons in this material system, the exciton comprises of
electron-hole states which are spread over a large momentum range
where a parabolic description of the dispersion is not valid. Since
TMD excitons are composed of electron-hole states around $\pm
K$-points where Berry curvature is finite, it is natural to ask if
Bloch-band geometry can alter the excitonic spectrum. While the role
of Berry curvature in determining transport properties of
non-interacting Bloch electrons is well established~\cite{Xiao10,
Neupert13}, its role in modifying Coulomb attraction leading to
bound-state formation has not been explicitly analyzed.

In this Letter, we analyze the excitonic signatures of the two
geometric invariants of Bloch bands -- Berry curvature and quantum
geometric tensor (QGT). We show that the Berry curvature is
primarily responsible for a reciprocal-space orbital Zeeman effect
leading to a finite splitting of the 2$p_x \pm i$2$p_y$ states. QGT
on the other hand, contributes to the 2$s$-2$p$ splitting, that is
similar to the Lamb shift. Our findings apply to Wannier-Mott
excitons in general, and are particularly relevant for TMD excitons
where optical spectroscopy can directly probe the above-mentioned
signatures of the Bloch-band geometry.

\paragraph{Exciton problem in momentum space \textemdash}
The exciton motion can be decomposed into the relative motion of electron and hole giving rise to hydrogen-like bound states, and the center-of-mass momentum ($K_{CM}$) resulting in an excitonic dispersion periodic in reciprocal lattice vector. As light emission and absorption takes places around $K_{CM} = 0$ due to negligible momentum of photons, in the following we will only consider excitons with zero center-of-mass momentum. Without loss of generality, we restrict ourselves to two-dimensional excitons for the rest of the discussion. The exciton wave function can be expressed as - $\sum_\mathbf{k} A_{\nu}(\mathbf{k})c_{c,\mathbf{k}}^{\dagger} c_{v,\mathbf{k}}|0\rangle = \sum_{\mathbf{k}} A_{\nu}(\mathbf{k})|\mathbf{k}\rangle$, with $c_c$ ($c_v$) being the annihilation operator for an electron in the conduction (valence) band, and $|0\rangle$ as the semiconductor vacuum with no excitations. Amplitudes $A_{\nu}(\mathbf{k})$s satisfy the eigenvalue equation $\sum_\mathbf{k'}\langle \mathbf{k} | \mathcal{H} + V |\mathbf{k'}\rangle A_{\nu}(\mathbf{k'}) = E A_{\nu}(\mathbf{k})$ where $\mathcal{H} = \sum_{\mathbf{k}}\mathcal{E}^v_{\mathbf{k}} c_{v,\mathbf{k}}^{\dagger} c_{v,\mathbf{k}} + \sum_{\mathbf{k}}\mathcal{E}^c_{\mathbf{k}}c_{c,\mathbf{k}}^{\dagger} c_{c,\mathbf{k}}$ is the single-particle Hamiltonian of the two-band semiconductor, and $V = \int d^2 r d^2 r' \Psi^\dagger (\mathbf{r}) \Psi^\dagger (\mathbf{r'})\frac{e^2}{\varepsilon |\mathbf{r} - \mathbf{r'}|} \Psi (\mathbf{r}) \Psi (\mathbf{r'})$ is the Coulomb interaction. Field operators are defined as - $\Psi(\mathbf{r}) = \frac{1}{\sqrt{S}}\sum_{n = v,c}\sum_{\mathbf{k}}u_{n,\mathbf{k}}(\mathbf{r})e^{i\mathbf{k}\cdot\mathbf{r}}c_{n,\mathbf{k}}$, where $u_{n,\mathbf{k}}(\mathbf{r})$ are the Bloch functions of the band with index $n = {c,v}$ and $S$ is the quantization area.

The matrix elements are given by $\langle \mathbf{k} | \mathcal{H} | \mathbf{k'}\rangle$ = $\delta_{\mathbf{k},\mathbf{k}'}\left(\mathcal{E}^c_{\mathbf{k}} - \mathcal{E}^v_{\mathbf{k}}\right)$, and $\langle \mathbf{k} | V | \mathbf{k}' \rangle$ = $-\frac{e^2}{S^2}\int d^2 r d^2 r' u^{*}_{c,\mathbf{k}}(\mathbf{r})u_{c,\mathbf{k}'}(\mathbf{r})  \frac{e^{i(\mathbf{k}' - \mathbf{k}) \cdot (\mathbf{r} - \mathbf{r}')}}{\varepsilon|\mathbf{r} - \mathbf{r}'|} u^{*}_{v,\mathbf{k}'}(\mathbf{r}')u_{v,\mathbf{k}}(\mathbf{r}')$ = $-\frac{1}{S}\frac{2\pi e^2}{\varepsilon}\frac{1}{|\mathbf{k} - \mathbf{k}'|} \langle u_{c,\mathbf{k}}|u_{c,\mathbf{k}'} \rangle \langle u_{v,\mathbf{k}'}| u_{v,\mathbf{k}} \rangle$, which is the direct part of Coulomb interaction. The long-range part of exchange interaction of $V$ vanishes for excitons with $K_{CM} = 0$. We neglect the short-range part of the exchange interaction since its magnitude is much smaller than the direct terms we analyze.

The eigenvalue equation for the exciton then becomes,
\begin{align}
(\Delta_{\mathbf{k}} - E_{\nu}) A_{\nu}(\mathbf{k}) -\frac{1}{S}\sum_{\mathbf{k}'}\frac{2\pi e^2}{\varepsilon|\mathbf{k} - \mathbf{k}'|} s^c_{\mathbf{k},\mathbf{k}'}s^v_{\mathbf{k}',\mathbf{k}}A_{\nu}(\mathbf{k}') = 0,
\label{ExcitonEq}
\end{align}
where the index $\nu$ = 1$s$, 2$s$, 2$p$, etc in analogy to hydrogenic orbitals, $\Delta_{\mathbf{k}} = \mathcal{E}^c_{\mathbf{k}}  - \mathcal{E}^v_{\mathbf{k}}$ and Bloch overlaps $s^{n}_{\mathbf{k},\mathbf{k}'} = \langle u_{n,\mathbf{k}}|u_{n,\mathbf{k}'} \rangle$. When the Bloch overlaps are unity, and the dispersion $\Delta_{\mathbf{k}}$ is parabolic, we recover the two-dimensional (2D) hydrogen atom solution.

%Guided by this observation, we choose Berry curvature, a gauge-invariant quantity dependent on geometry of Bloch bands, as the parameter for perturbative expansion of $s^n_{\mathbf{k},\mathbf{k}'}$.
%In order to illustrate the effect of Bloch overlaps, we treat them as perturbation to the 2d hydrogen atom problem.

\paragraph{Bloch part: General arguments \textemdash}
%Having demonstrated the role of constant Berry curvature in modification of exciton spectra, we turn to the general case where Berry curvature varies in $k$-space as dictated by crystal symmetry.
It is clear from Eq. (\ref{ExcitonEq}) that Bloch overlaps $s^n_{\mathbf{k},\mathbf{k}'}$ enforce the symmetry of the crystal on Coulomb attraction between electron-hole pair. The radial symmetry of hydrogen atom is lowered when $s^n_{\mathbf{k},\mathbf{k}'}$ deviate significantly from unity thereby necessarily making the excitonic spectrum non-hydrogenic. This should be contrasted with a deviation arising from the non-parabolicity of single-particle states. Note that Bloch overlaps as defined above are gauge-dependent unlike the eigen-energies $E_{\nu}$. They need not change with band dispersion but depend on how the Bloch functions are arranged in $\mathbf{k}$-space or the geometry of Bloch bands $u_{n,\mathbf{k}}$ over the Brillouin zone. Guided by this observation, below we express $s^n_{\mathbf{k},\mathbf{k}'}$ in terms of geometric invariants of Bloch bands such as Berry curvature. We first note that the Bloch overlap $s^n_{\mathbf{k},\mathbf{k}'}$ for $\mathbf{k'} \sim \mathbf{k} + d\mathbf{k}$ can be expressed as -
\begin{eqnarray}
\label{infOverlap}
s^n_{\mathbf{k},\mathbf{k}+d\mathbf{k}} = &1& + \langle u_n(\mathbf{k})| \partial_{k_i}|u_n(\mathbf{k})\rangle dk_i \\
&+& \frac{1}{2}\langle u_n(\mathbf{k})| \partial_{k_i}\partial_{k_j}|u_n(\mathbf{k})\rangle dk_i dk_j + \ldots \nonumber
\end{eqnarray}

The first order term in $d\mathbf{k}$ is the Berry connection $i\mathcal{A}_{i}$ which is related to Berry curvature as $\mathbf{\Omega}(\mathbf{k})$ = $\nabla \times \bf{\mathcal{A}}(\mathbf{k})$~\cite{Xiao10}. Eq. (\ref{infOverlap}) is not gauge-invariant in that it does not transform as a tensor under the transformation $\tilde{u}_n(\mathbf{k})$ = $u_n(\mathbf{k})e^{i\alpha(\mathbf{k})}$. However, if one chooses a closed path in $\mathbf{k}$-space, arbitrary phases which arise under the $U(1)$ gauge-transformation, mutually cancel each other to give gauge-invariant quantities. Indeed, only such closed-loop terms appear in the characteristic polynomial of the eigenvalue problem in Eq. (\ref{ExcitonEq}). It can be shown that up to second order in $d\mathbf{k}$ (see Appendix A.) -
\begin{align}
\langle u_{\mathbf{k}_1}|u_{\mathbf{k}_2}\rangle \ldots \langle u_{\mathbf{k}_{N-1}}|u_{\mathbf{k}_N}\rangle \sim e^{\left({i\oint \mathbf{\mathcal{A}}\cdot d\mathbf{k}} - \frac{1}{2} \int g_{ij} dk_i dk_j \right)},
\label{Blochoverlap}
\end{align}
for a closed path such that $u_{\mathbf{k}_N}$ = $u_{\mathbf{k}_1}$. The first exponential on RHS is nothing but the Berry phase of the closed path while the second term in the exponential is the squared ``length" of the path defined in terms of the quantum geometric tensor $g_{ij}$ = $\mathrm{Re}\left[\langle \partial_{k_i} u(\mathbf{k})|\partial_{k_j} u(\mathbf{k})\rangle \right] - \mathcal{A}_i \mathcal{A}_j$. QGT, also referred to as Fubini-Study metric, is a gauge-invariant quantity corresponding to the second order derivative in Eq. (\ref{infOverlap}) which measures infinitesimal distance between Bloch states parametrized by $\mathbf{k}$~\cite{Provost80}. Thus, the gauge-invariant Bloch overlaps over a closed loop can be expressed solely in terms of geometric quantities characterizing the Bloch bands.
It is instructive to study the effect of Bloch overlaps perturbatively for which we consider an infinitesimal loop, when all $\mathbf{k}$s are close to each other. Eq. (\ref{Blochoverlap}) then becomes -
\begin{align}
1 + i \mathbf{\Omega} \cdot d\mathbf{S}_k - \frac{1}{2} g_{ij} dk_i dk_j + \ldots,
\label{infinitesimalGeneral}
\end{align}
where we have expressed $\oint \mathbf{\mathcal{A}}\cdot d\mathbf{k}$ in terms of Berry curvature as $ \int \mathbf{\Omega} \cdot d\mathbf{S}_k$ using Stoke's theorem. It should be noted that the imaginary part of the above expression is proportional to Berry curvature which is an antisymmetric quantity whereas the real part is proportional to QGT, a symmetric quantity. When calculating Bloch overlaps for conduction and valence bands, the difference (sum) of the Berry curvatures (QGTs) of the two bands appears.
%The above expression is a generalization of Eq. (\ref{BlochPert}) beyond the gapped graphene model, as long as the extent of exciton in $\mathbf{k}$-space is small.

\paragraph{Illustrative model: Gapped graphene \textemdash}
Having established a connection between Bloch overlaps and geometry of Bloch bands, we now illustrate how Berry curvature and QGT effect the exciton spectrum using a toy-model. We choose a simple two-band model of graphene with a band gap at $\pm K$-points for which Bloch overlaps $s^n_{\mathbf{k},\mathbf{k}'}$ can be analytically obtained. The Hamiltonian can be written in Pauli basis of the two bands as $H(\mathbf{k}) = (atk_x, at\tau_v k_y, \Delta_0)$ where $a$ is the lattice constant, $t$ denotes the hopping energy, and $\tau_v = \pm1$ is the valley index. The presence of inversion symmetry breaking band gap $\Delta_0$ results in equal but opposite Berry curvature near the valleys at $\pm K$-points~\cite{Xiao07, Yao08}. Furthermore, the magnitude of $\Delta_0$ is chosen to be large enough such that the magnitude of Berry curvature at the $\pm K$-points $|\Omega_0| = a^2t^2/\Delta_0^2$ is small and can be taken to be constant in the $\mathbf{k}$ region where the exciton wave function extends i.e., $\Omega_0 |\mathbf{k}|^2 \ll 1$ for $|\mathbf{k}| \in \delta k \sim 1/a_B$ around the $\pm K$-points.

Under these assumptions,
\begin{align}
s^c_{\mathbf{k},\mathbf{k}'}s^v_{\mathbf{k}',\mathbf{k}} = 1 + \frac{|\Omega_0|}{2}\left(i\tau_v (\mathbf{k'} \times \mathbf{k}) -\frac{1}{2}|\mathbf{k} -\mathbf{k'}|^2 \right) + \ldots,
\label{BlochPert}
\end{align}
up to first order in $|\Omega_0|$. Thus, $|\Omega_0|$ serves as a small parameter for perturbative treatment of the Bloch part. Note that Eq. (\ref{BlochPert}) is analogous to Eq. (\ref{infinitesimalGeneral}) for the present model. In particular, the imaginary, antisymmetric part (opposite in the two valleys) corresponds to the Berry curvature while the real symmetric part to QGT. In the present model, both Berry curvature and QGT are proportional to $|\Omega_0|$ = $a^2 t^2 /\Delta_0 ^2$. Plugging Eq. (\ref{BlochPert}) in (\ref{ExcitonEq}), one can write the exciton Hamiltonian of Eq. (\ref{ExcitonEq}) perturbatively as $H_{\mathrm{ex}} = H^{\mathrm{H}} + V^{I}$, where $H^{\mathrm{H}}$ is the Hamiltonian without the Bloch part describing the 2D hydrogen atom while $V^{I}$ is the perturbative term due to Bloch overlaps which reads as -
\begin{align}
V^{I} = \frac{|\Omega_0|}{2}\frac{2\pi e^2}{S \varepsilon}\left(\frac{1}{2}|\mathbf{k} -\mathbf{k'}| - i\tau_v \frac{(\mathbf{k'} \times \mathbf{k})}{|\mathbf{k} -\mathbf{k}' |} \right).
\end{align}
We decompose $V^{I}$ into a real, symmetric part, $V^I_{S}$ and an imaginary, antisymmetric part, $V^{I}_{AS}$, making $V^{I}$ Hermitian, as expected.

In order to determine the new eigenvalues and eigenstates of the perturbed Hamiltonian, one needs calculate the matrix elements of $V_I$ in the basis of 2D hydrogenic levels. We first consider the effect of $V^{I}$ on $2p_x$ and $2p_y$ states which are degenerate in 2D hydrogen atom.
Due to the symmetry of $V^{I}_{AS}$ under the exchange of $\mathbf{k}$ and $\mathbf{k}'$, its diagonal matrix elements vanish, while the off-diagonal matrix elements are finite. Likewise, $V^{I}_{S}$ has only diagonal matrix elements which are non-zero.

The new eigenstates are obtained by diagonalizing the following matrix -
\begin{align}
\begin{pmatrix}
E_{2p}^{H} + \langle\psi_{2p_x}|V^I_S| \psi_{2p_x}\rangle & \langle\psi_{2p_x}|V^I_{AS}| \psi_{2p_y}\rangle \\ \\
\langle\psi_{2p_y}|V^I_{AS}| \psi_{2p_x}\rangle & E_{2p}^H + \langle\psi_{2p_y}|V^I_S| \psi_{2p_y}\rangle
\end{pmatrix}
\label{2pmatrix}
\end{align}
It is clear that the off-diagonal matrix elements arising from the antisymmetric part of the $V^{I}$ will mix the $2p_x$ and $2p_y$ states into symmetric and antisymmetric combinations $|\psi_{2p_{\pm}}\rangle = \frac{1}{\sqrt{2}} \left(|\psi_{2p{_x}}\rangle \pm i|\psi_{2p_{y}}\rangle\right)$ and causing them to split. On the other hand, the diagonal matrix elements due to the symmetric part of $V^{I}$ will cause rigid shift in energies of $|\psi_{2p_\pm}\rangle$ by the same amount.

The above scenario is an analog of orbital Zeeman effect where degeneracy of $2p$ states of a hydrogen atom is lifted in the presence of a constant magnetic field due to coupling to angular momentum $\l$, an antisymmetric quantity. Berry curvature is the momentum space analog of magnetic field and the splitting here can be thought of as a ``momentum-space orbital Zeeman effect". In general, the degeneracy of all $l \neq 0$ states of 2D hydrogen atom will be lifted due to Berry curvature as in the case of orbital Zeeman effect. Due to the appearance of valley index $\tau_v$ in $V^{I}_{AS}$, the splitting is opposite in the two valleys as required by time-reversal symmetry. From Eq. (\ref{infinitesimalGeneral}), we can conclude that when Berry curvature is small but not constant, it is the Berry flux through the exciton wave function which determines the splitting. We note the similarity of our findings with that of Ref.~\cite{Garate11} where excitons on the surface of topological insulators with explicitly broken time-reversal symmetry were considered.

\begin{figure}
 \includegraphics[width=1.0\linewidth]{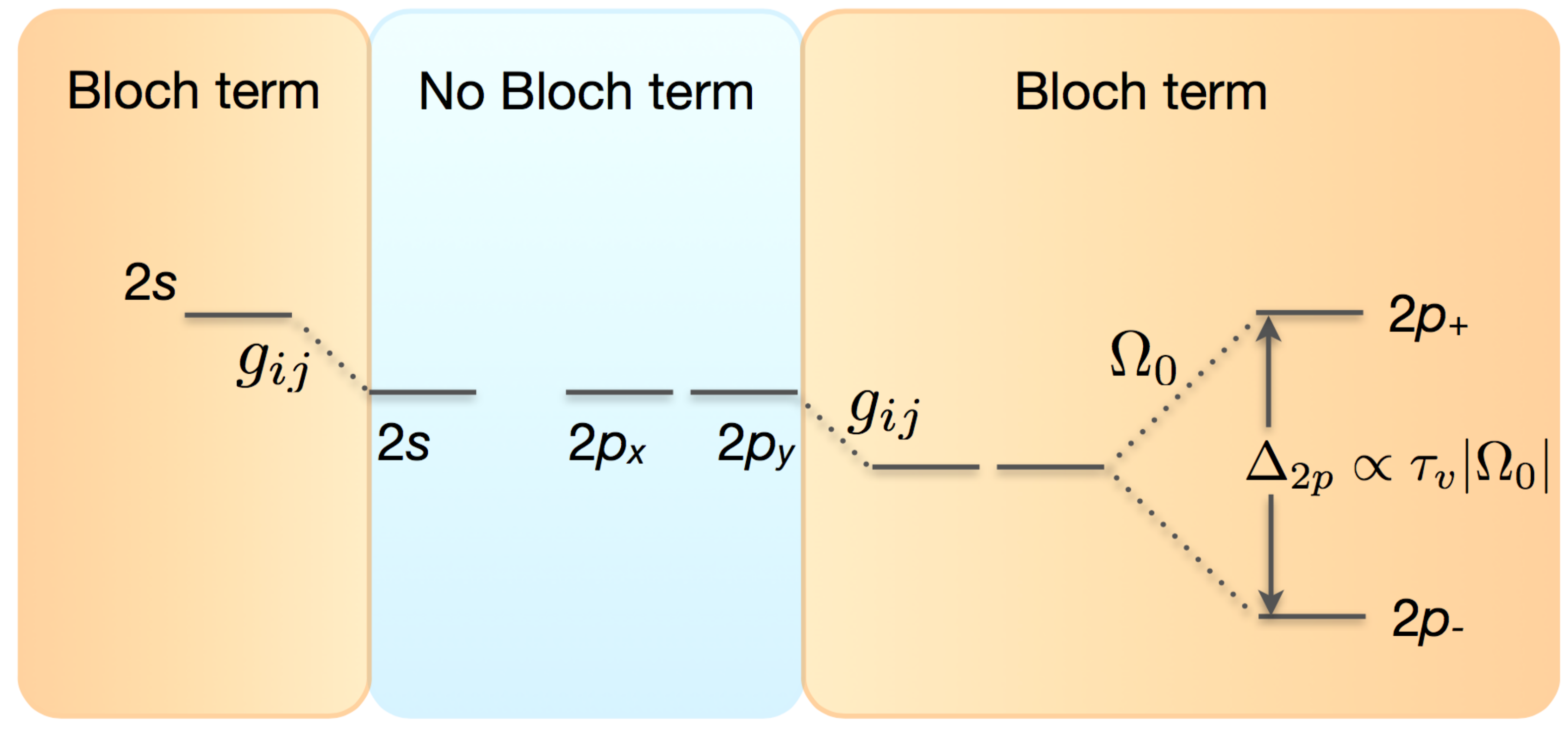}
  \caption{ (Color online) A sketch of energy levels of the gapped-graphene model under assumptions of small, constant Berry curvature ($\Omega_0$) and large exciton Bohr-radius. In absence of Bloch term in the numerical calculations, $n$ = 2 states remain degenerate like hydrogen atom. Upon the inclusion of Bloch term, 2$p$ states mix and split due to $\Omega_0$ much like the orbital Zeeman effect. The shift in energy levels arises from quantum geometric tensor ($g_{ij}$) and is analogous to Lamb shift in hydrogen atom.
 }
 \label{fig:gappedGraphene}
\end{figure}

Under the assumption of small, constant $\Omega_0$ and $\mathbf{k}$ close to $K$-point, we numerically evaluate the matrix elements in Eq. (\ref{2pmatrix}) to be $\langle\psi_{2p_x}|V^I_{AS}| \psi_{2p_y}\rangle$ = $-i\tau_v |\Omega_0|c/2$ and $\langle\psi_{2p_x}|V^I_{S}| \psi_{2p_x}\rangle$  $\sim$ $-|\Omega_0|c/4$, where $c >$ 0 is a constant. The energies of $2p_{\pm}$ states are obtained to be $E_{2p_{\pm}}$ = $E_{2p} - |\Omega_0|c/4 \pm |\Omega_0|c/2$. The splitting between the $2p$ states is then $\Delta_{2p} \propto \tau_v |\Omega_0|$. For $s$-states, only the symmetric part $V^{I}_{S}$ survives due to symmetry reasons, leading to a blue-shift in energy. We find that the shift of $2s$ states is $\sim |\Omega_0|c/4$, making it almost degenerate with $2p_{+}$ state under our assumptions. Figure~\ref{fig:gappedGraphene}, shows a schematic energy-level diagram of the model with and without the Bloch perturbation. Upon setting the Bloch overlap to be unity, we find that the $2p$ states remain degenerate resembling unmixed $2p_x$ and $2p_y$ orbitals thereby confirming the role of Bloch overlap in splitting.

One can also consider a two-band model with identically zero Berry curvature to confirm that the mixing of $2p_x$ and $2p_y$ states stems from Berry curvature. If $\left(h_x\left(k_x, k_y\right), h_y\left(k_x, k_y\right), \Delta_0\right)$ is the Hamiltonian in Pauli matrix basis, then choosing $h_x\left(k_x,k_y\right)$ $\propto$ $h_y\left(k_x,k_y\right)$ gives vanishing Berry curvature as it is proportional to $\partial h_x \times \partial h_y$. Indeed, in numerical calculations we find that the $2p$ wave functions for such a model remain unmixed.

Next, we comment on the role of QGT in determining the exciton spectra. QGT can be rewritten as -
\begin{eqnarray}
g_{ij} &=& \mathrm{Re}\left[ \langle \partial_{k_i}u|\partial_{k_j u} \rangle\right] - \langle \partial_{k_i}u | u \rangle \langle u |\partial_{k_j}  u \rangle \\
 &=& \langle X_i X_j \rangle - \langle X_i \rangle \langle X_j \rangle, \nonumber
\end{eqnarray}
where, the operator $X$ is the generator of translation in the $\mathbf{k}$-space. QGT then measures the quantum fluctuations of $X$~\cite{Provost80}. In the present case, $X$ corresponds to a spread in ``relative position" of electron and hole. This is not unlike the case of Lamb-shift in hydrogen atom where vacuum fluctuations of the electromagnetic field smear electron's position, in turn changing its potential energy and resulting in a measurable shift of its energy~\cite{Welton48}. In other words, QGT can be thought of as ``vacuum fluctuations" of the effective $U(1)$ gauge field present in the $\mathbf{k}$-space, and the corresponding shifts as an analog of the Lamb shift. Even when Berry curvature vanishes identically, the effect of QGT must still remain. It is noteworthy that the magnitude of this Lamb-like shift can be relatively large compared to the hydrogenic Lamb shift; this observation can be thought of as a consequence of the fact that the effective fine structure constant of the gapped graphene model, $\alpha$ = $e^2/\hbar v$ is on the the order of unity. Thus, we identify another physically relevant consequence of QGT which has been previously shown to be behind such varied phenomena as  electric polarization in insulators~\cite{Marzari97}, current noise~\cite{Neupert13}, and even quantum phase transitions~\cite{Ma10, Zanardi07}
 %%% Fig 1 : energy diagram of gapped graphene model
\paragraph{TMD excitons \textemdash}
In the following, we investigate how the predictions of the preceding discussion apply to TMD excitons. We first note that the experimentally observed deviation from a hydrogenic series arises due to a combination of non-local dielectric screening, non-parabolic dispersion of bands, and the effect of the Bloch part. The assumption of constant and small Berry curvature no longer holds due to the large spread of exciton wave function in $\mathbf{k}$-space. In addition, unlike other material system such as GaAs  where the lowest energy exciton is made from electron-hole states near the $\Gamma$-point ($\mathbf{k}$= 0), lowest energy excitons in TMDs are made from $\pm K$-point electron-hole pairs where there is non-zero Berry curvature. Thus, we expect a mixing and splitting of $2p$-states in addition to energy shifts due to QGT.

We solve for the eigenvalue problem in Eq. (\ref{ExcitonEq}) using a three-band, next-nearest neighbor model, for MoS$_2$ and WSe$_2$,  which captures the band dispersion of conduction and valence band throughout the Brillouin zone~\cite{Liu13}. We discretize the Brillouin zone into a grid of 136 $\times$ 136 $\mathbf{k}$-points and assume an air-suspended sample with a non-local dielectric screening length $r^{*} \sim$ 15 \AA,~corresponding to a binding energy of $\sim$ 400 meV for the $1s$-state which, is estimated to be in the range of 300 - 700 meV~\cite{Qiu13, Chernikov14, Wang15}. The Coulomb part of the Hamiltonian is calculated such that $|\mathbf{k}-\mathbf{k}'|$ is always restricted to the first Brillouin zone.
\begin{figure}
 \includegraphics[width=1.0\linewidth]{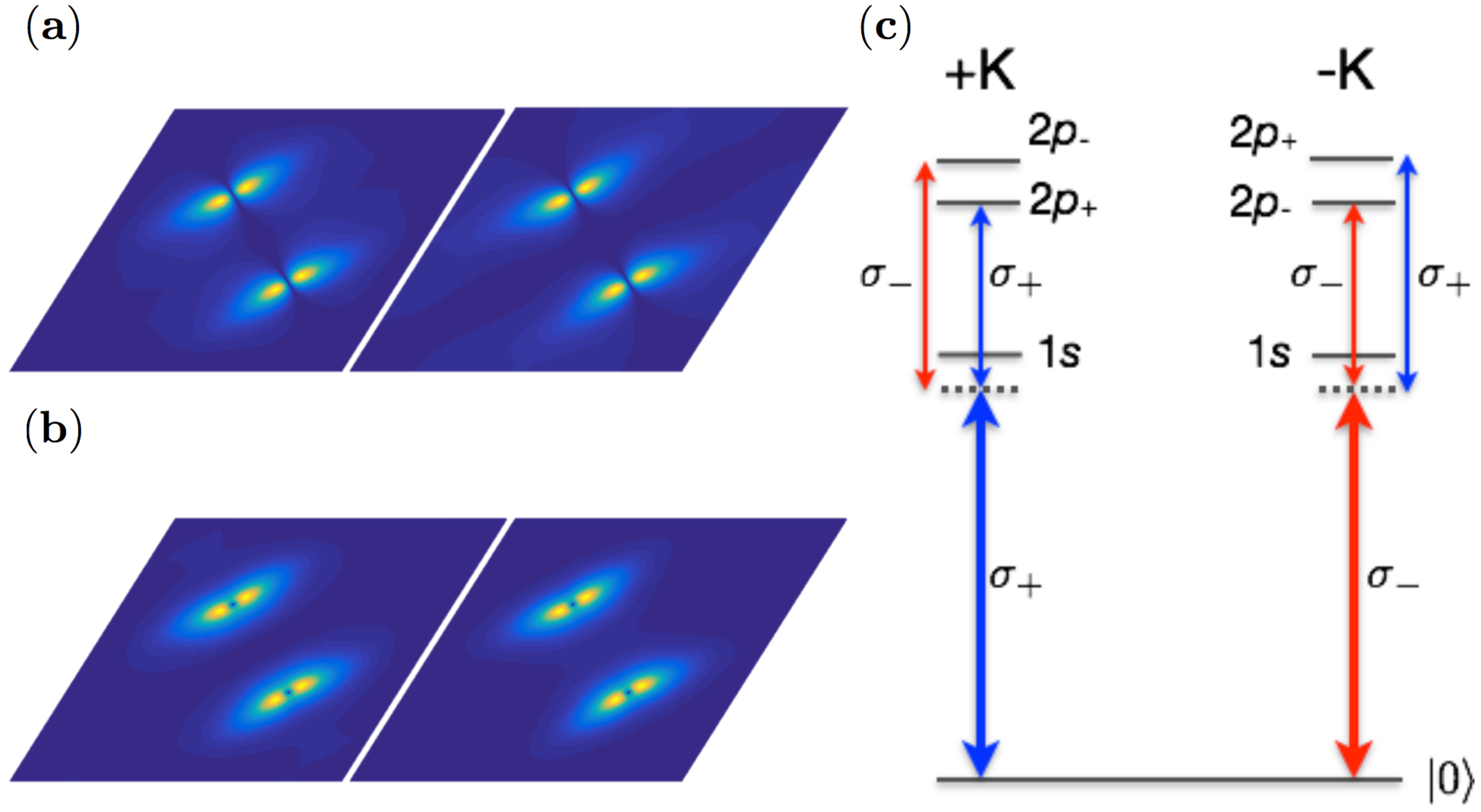}
  \caption{(Color online) Calculated squared amplitude of 2$p$ wave functions in reciprocal space for MoS$_2$ in ({\bf{a}}) absence and ({\bf{b}}) presence of Bloch overlap. The Brillouin zone is chosen such that $\Gamma$-points lie on the four vertices. The exciton wave function extends near $\pm K$ points. Without the Bloch perturbation, the wave functions are degenerate and resemble 2$p_x$ and 2$p_y$ wave functions. Bloch perturbation mixes the states which resemble 2$p_{\pm}$ states and have an energy splitting of about $\sim$ 10 meV. {\bf (c)} A sketch of energy levels with a scheme to optically determine the predicted splitting of 2$p$-states using two-photon resonance spectroscopy.}
 \label{fig:WSe2}
\end{figure}
	
Figure~\ref{fig:WSe2} (a)-(b) show the $2p$ wave functions for MoS$_2$ exciton with and without the Bloch part. Just like in the case of gapped graphene, $2p$ states remain unmixed without the inclusion of Bloch part in calculations. With the Bloch term included, we obtain a $2p$ splitting $\sim$ 10 meV ($\sim$ 14 meV) for MoS$_2$ (WSe$_2$) which is consistent with the recently reported splitting of $\sim$ 25 meV in MoS$_2$~\cite{Wu15}. Given the relatively large splitting, it should be possible to detect it experimentally using two-photon, polarization resolved, optical spectroscopy involving a near-infrared laser slightly detuned from the 1$s$ exciton and a mid-infrared laser as shown in Fig.~\ref{fig:WSe2}(b). When the two lasers are co-circularly (counter-circularly) polarized, lower (higher) energy 2$p$ state will be excited in a two-photon resonance leading to an enhancement in photoluminescence from the 1$s$ state. We note that even in the absence of Bloch part, trigonal warping of the dispersion can lead to a degeneracy lifting between $2p_x$ and $2p_y$ states, however, they remain unmixed and the magnitude of splitting is much smaller than the Berry curvature induced splitting. We also note that the level diagram in Fig.~\ref{fig:WSe2}(b) is strictly true for $K_{CM}$ = 0, as the long-range part of exchange interaction will mix the states in opposite valleys for finite $K_{CM}$~\cite{Yu14}. 
\paragraph{Conclusions \textemdash}
The central role played by the Berry curvature in determining the transport properties of non-interacting Bloch electrons, leading to anomalous, valley and spin Hall effects, is well established. Our results on the other hand, unequivocally demonstrate that the spectrum of Coulomb-correlated two-particle bound states exhibit observable signatures of Berry curvature and QGT. This motivates the question whether many-body optical excitations of semiconductors, such as trions in a 2D electron system, are also influenced by effective gauge-fields in solids arising due to non-trivial geometry of Bloch-bands.

\acknowledgements We thank Charles Grenier for fruitful discussions. This work is supported by NCCR Quantum Science and Technology (NCCR QSIT), research instrument of the Swiss National Science Foundation (SNSF).

%\begin{figure}[h]
%  \centering
%  % Requires \usepackage{graphicx}
%  \includegraphics[width=3in]{singleshotfig4}\\
%  \caption{(a) Quantum jumps in continuous readout. The power of laser used for exciting the diagonal transition is weaker for the upper panel ($P_1$) and stronger for  the lower panel ($P_2 > P_1$). (b) Same data as (a), lower panel, but the counts are stored in 1$\mu s$ bins. (c)  Second order correlation function $g^{(2)}(t)$ (black trace) and waiting time distribution (red trace) for the data in the upper panel of (a). The dashed line shows the spin lifetime.  Inset: the normalized waiting time function for the upper panel  (red line) and the lower panel (blue line) in (a). Two exponential decays are observed.  The fitted decay times for the red line are $t_{wait}=349 \pm 5 ns$ and $t_{repump}= 12.1\pm 0.3 \mu s$ and for the blue line $t_{wait}=371 \pm 5 ns$ and $t_{repump}= 6.32\pm 0.07 \mu s$ }\label{1}
%\end{figure}

%%%% References

%
 %\bibliography{ExcitonBerryCurvature_arXiv}

\begin{thebibliography}{24}%
\makeatletter
\providecommand \@ifxundefined [1]{%
 \@ifx{#1\undefined}
}%
\providecommand \@ifnum [1]{%
 \ifnum #1\expandafter \@firstoftwo
 \else \expandafter \@secondoftwo
 \fi
}%
\providecommand \@ifx [1]{%
 \ifx #1\expandafter \@firstoftwo
 \else \expandafter \@secondoftwo
 \fi
}%
\providecommand \natexlab [1]{#1}%
\providecommand \enquote  [1]{``#1''}%
\providecommand \bibnamefont  [1]{#1}%
\providecommand \bibfnamefont [1]{#1}%
\providecommand \citenamefont [1]{#1}%
\providecommand \href@noop [0]{\@secondoftwo}%
\providecommand \href [0]{\begingroup \@sanitize@url \@href}%
\providecommand \@href[1]{\@@startlink{#1}\@@href}%
\providecommand \@@href[1]{\endgroup#1\@@endlink}%
\providecommand \@sanitize@url [0]{\catcode `\\12\catcode `\$12\catcode
  `\&12\catcode `\#12\catcode `\^12\catcode `\_12\catcode `\%12\relax}%
\providecommand \@@startlink[1]{}%
\providecommand \@@endlink[0]{}%
\providecommand \url  [0]{\begingroup\@sanitize@url \@url }%
\providecommand \@url [1]{\endgroup\@href {#1}{\urlprefix }}%
\providecommand \urlprefix  [0]{URL }%
\providecommand \Eprint [0]{\href }%
\providecommand \doibase [0]{http://dx.doi.org/}%
\providecommand \selectlanguage [0]{\@gobble}%
\providecommand \bibinfo  [0]{\@secondoftwo}%
\providecommand \bibfield  [0]{\@secondoftwo}%
\providecommand \translation [1]{[#1]}%
\providecommand \BibitemOpen [0]{}%
\providecommand \bibitemStop [0]{}%
\providecommand \bibitemNoStop [0]{.\EOS\space}%
\providecommand \EOS [0]{\spacefactor3000\relax}%
\providecommand \BibitemShut  [1]{\csname bibitem#1\endcsname}%
\let\auto@bib@innerbib\@empty
%</preamble>
\bibitem [{\citenamefont {Byrnes}\ \emph {et~al.}(2014)\citenamefont {Byrnes},
  \citenamefont {Kim},\ and\ \citenamefont {Yamamoto}}]{Byrnes14}%
  \BibitemOpen
  \bibfield  {author} {\bibinfo {author} {\bibfnamefont {T.}~\bibnamefont
  {Byrnes}}, \bibinfo {author} {\bibfnamefont {N.~Y.}\ \bibnamefont {Kim}}, \
  and\ \bibinfo {author} {\bibfnamefont {Y.}~\bibnamefont {Yamamoto}},\ }\href
  {http://dx.doi.org/10.1038/nphys3143} {\bibfield  {journal} {\bibinfo
  {journal} {Nat Phys}\ }\textbf {\bibinfo {volume} {10}},\ \bibinfo {pages}
  {803} (\bibinfo {year} {2014})}\BibitemShut {NoStop}%
\bibitem [{\citenamefont {Smolka}\ \emph {et~al.}(2014)\citenamefont {Smolka},
  \citenamefont {Wuester}, \citenamefont {Haupt}, \citenamefont {Faelt},
  \citenamefont {Wegscheider},\ and\ \citenamefont {Imamoglu}}]{Smolka14}%
  \BibitemOpen
  \bibfield  {author} {\bibinfo {author} {\bibfnamefont {S.}~\bibnamefont
  {Smolka}}, \bibinfo {author} {\bibfnamefont {W.}~\bibnamefont {Wuester}},
  \bibinfo {author} {\bibfnamefont {F.}~\bibnamefont {Haupt}}, \bibinfo
  {author} {\bibfnamefont {S.}~\bibnamefont {Faelt}}, \bibinfo {author}
  {\bibfnamefont {W.}~\bibnamefont {Wegscheider}}, \ and\ \bibinfo {author}
  {\bibfnamefont {A.}~\bibnamefont {Imamoglu}},\ }\href {\doibase
  10.1126/science.1258595} {\bibfield  {journal} {\bibinfo  {journal}
  {Science}\ }\textbf {\bibinfo {volume} {346}},\ \bibinfo {pages} {332}
  (\bibinfo {year} {2014})}\BibitemShut {NoStop}%
\bibitem [{\citenamefont {Kazimierczuk}\ \emph {et~al.}(2014)\citenamefont
  {Kazimierczuk}, \citenamefont {Frohlich}, \citenamefont {Scheel},
  \citenamefont {Stolz},\ and\ \citenamefont {Bayer}}]{Kazimierczuk14}%
  \BibitemOpen
  \bibfield  {author} {\bibinfo {author} {\bibfnamefont {T.}~\bibnamefont
  {Kazimierczuk}}, \bibinfo {author} {\bibfnamefont {D.}~\bibnamefont
  {Frohlich}}, \bibinfo {author} {\bibfnamefont {S.}~\bibnamefont {Scheel}},
  \bibinfo {author} {\bibfnamefont {H.}~\bibnamefont {Stolz}}, \ and\ \bibinfo
  {author} {\bibfnamefont {M.}~\bibnamefont {Bayer}},\ }\href
  {http://dx.doi.org/10.1038/nature13832} {\bibfield  {journal} {\bibinfo
  {journal} {Nature}\ }\textbf {\bibinfo {volume} {514}},\ \bibinfo {pages}
  {343} (\bibinfo {year} {2014})}\BibitemShut {NoStop}%
\bibitem [{\citenamefont {Ye}\ \emph {et~al.}(2014)\citenamefont {Ye},
  \citenamefont {Cao}, \citenamefont {O'Brien}, \citenamefont {Zhu},
  \citenamefont {Yin}, \citenamefont {Wang}, \citenamefont {Louie},\ and\
  \citenamefont {Zhang}}]{Ye14}%
  \BibitemOpen
  \bibfield  {author} {\bibinfo {author} {\bibfnamefont {Z.}~\bibnamefont
  {Ye}}, \bibinfo {author} {\bibfnamefont {T.}~\bibnamefont {Cao}}, \bibinfo
  {author} {\bibfnamefont {K.}~\bibnamefont {O'Brien}}, \bibinfo {author}
  {\bibfnamefont {H.}~\bibnamefont {Zhu}}, \bibinfo {author} {\bibfnamefont
  {X.}~\bibnamefont {Yin}}, \bibinfo {author} {\bibfnamefont {Y.}~\bibnamefont
  {Wang}}, \bibinfo {author} {\bibfnamefont {S.~G.}\ \bibnamefont {Louie}}, \
  and\ \bibinfo {author} {\bibfnamefont {X.}~\bibnamefont {Zhang}},\ }\href
  {http://dx.doi.org/10.1038/nature13734} {\bibfield  {journal} {\bibinfo
  {journal} {Nature}\ }\textbf {\bibinfo {volume} {513}},\ \bibinfo {pages}
  {214} (\bibinfo {year} {2014})}\BibitemShut {NoStop}%
\bibitem [{\citenamefont {He}\ \emph {et~al.}(2014)\citenamefont {He},
  \citenamefont {Kumar}, \citenamefont {Zhao}, \citenamefont {Wang},
  \citenamefont {Mak}, \citenamefont {Zhao},\ and\ \citenamefont
  {Shan}}]{He14}%
  \BibitemOpen
  \bibfield  {author} {\bibinfo {author} {\bibfnamefont {K.}~\bibnamefont
  {He}}, \bibinfo {author} {\bibfnamefont {N.}~\bibnamefont {Kumar}}, \bibinfo
  {author} {\bibfnamefont {L.}~\bibnamefont {Zhao}}, \bibinfo {author}
  {\bibfnamefont {Z.}~\bibnamefont {Wang}}, \bibinfo {author} {\bibfnamefont
  {K.~F.}\ \bibnamefont {Mak}}, \bibinfo {author} {\bibfnamefont
  {H.}~\bibnamefont {Zhao}}, \ and\ \bibinfo {author} {\bibfnamefont
  {J.}~\bibnamefont {Shan}},\ }\href {\doibase 10.1103/PhysRevLett.113.026803}
  {\bibfield  {journal} {\bibinfo  {journal} {Phys. Rev. Lett.}\ }\textbf
  {\bibinfo {volume} {113}},\ \bibinfo {pages} {026803} (\bibinfo {year}
  {2014})}\BibitemShut {NoStop}%
\bibitem [{\citenamefont {Chernikov}\ \emph {et~al.}(2014)\citenamefont
  {Chernikov}, \citenamefont {Berkelbach}, \citenamefont {Hill}, \citenamefont
  {Rigosi}, \citenamefont {Li}, \citenamefont {Aslan}, \citenamefont
  {Reichman}, \citenamefont {Hybertsen},\ and\ \citenamefont
  {Heinz}}]{Chernikov14}%
  \BibitemOpen
  \bibfield  {author} {\bibinfo {author} {\bibfnamefont {A.}~\bibnamefont
  {Chernikov}}, \bibinfo {author} {\bibfnamefont {T.~C.}\ \bibnamefont
  {Berkelbach}}, \bibinfo {author} {\bibfnamefont {H.~M.}\ \bibnamefont
  {Hill}}, \bibinfo {author} {\bibfnamefont {A.}~\bibnamefont {Rigosi}},
  \bibinfo {author} {\bibfnamefont {Y.}~\bibnamefont {Li}}, \bibinfo {author}
  {\bibfnamefont {O.~B.}\ \bibnamefont {Aslan}}, \bibinfo {author}
  {\bibfnamefont {D.~R.}\ \bibnamefont {Reichman}}, \bibinfo {author}
  {\bibfnamefont {M.~S.}\ \bibnamefont {Hybertsen}}, \ and\ \bibinfo {author}
  {\bibfnamefont {T.~F.}\ \bibnamefont {Heinz}},\ }\href {\doibase
  10.1103/PhysRevLett.113.076802} {\bibfield  {journal} {\bibinfo  {journal}
  {Phys. Rev. Lett.}\ }\textbf {\bibinfo {volume} {113}},\ \bibinfo {pages}
  {076802} (\bibinfo {year} {2014})}\BibitemShut {NoStop}%
\bibitem [{\citenamefont {Wang}\ \emph {et~al.}(2015)\citenamefont {Wang},
  \citenamefont {Marie}, \citenamefont {Gerber}, \citenamefont {Amand},
  \citenamefont {Lagarde}, \citenamefont {Bouet}, \citenamefont {Vidal},
  \citenamefont {Balocchi},\ and\ \citenamefont {Urbaszek}}]{Wang15}%
  \BibitemOpen
  \bibfield  {author} {\bibinfo {author} {\bibfnamefont {G.}~\bibnamefont
  {Wang}}, \bibinfo {author} {\bibfnamefont {X.}~\bibnamefont {Marie}},
  \bibinfo {author} {\bibfnamefont {I.}~\bibnamefont {Gerber}}, \bibinfo
  {author} {\bibfnamefont {T.}~\bibnamefont {Amand}}, \bibinfo {author}
  {\bibfnamefont {D.}~\bibnamefont {Lagarde}}, \bibinfo {author} {\bibfnamefont
  {L.}~\bibnamefont {Bouet}}, \bibinfo {author} {\bibfnamefont
  {M.}~\bibnamefont {Vidal}}, \bibinfo {author} {\bibfnamefont
  {A.}~\bibnamefont {Balocchi}}, \ and\ \bibinfo {author} {\bibfnamefont
  {B.}~\bibnamefont {Urbaszek}},\ }\href {\doibase
  10.1103/PhysRevLett.114.097403} {\bibfield  {journal} {\bibinfo  {journal}
  {Phys. Rev. Lett.}\ }\textbf {\bibinfo {volume} {114}},\ \bibinfo {pages}
  {097403} (\bibinfo {year} {2015})}\BibitemShut {NoStop}%
\bibitem [{\citenamefont {Ugeda}\ \emph {et~al.}(2014)\citenamefont {Ugeda},
  \citenamefont {Bradley}, \citenamefont {Shi}, \citenamefont {da~Jornada},
  \citenamefont {Zhang}, \citenamefont {Qiu}, \citenamefont {Ruan},
  \citenamefont {Mo}, \citenamefont {Hussain}, \citenamefont {Shen},
  \citenamefont {Wang}, \citenamefont {Louie},\ and\ \citenamefont
  {Crommie}}]{Ugeda14}%
  \BibitemOpen
  \bibfield  {author} {\bibinfo {author} {\bibfnamefont {M.~M.}\ \bibnamefont
  {Ugeda}}, \bibinfo {author} {\bibfnamefont {A.~J.}\ \bibnamefont {Bradley}},
  \bibinfo {author} {\bibfnamefont {S.-F.}\ \bibnamefont {Shi}}, \bibinfo
  {author} {\bibfnamefont {F.~H.}\ \bibnamefont {da~Jornada}}, \bibinfo
  {author} {\bibfnamefont {Y.}~\bibnamefont {Zhang}}, \bibinfo {author}
  {\bibfnamefont {D.~Y.}\ \bibnamefont {Qiu}}, \bibinfo {author} {\bibfnamefont
  {W.}~\bibnamefont {Ruan}}, \bibinfo {author} {\bibfnamefont {S.-K.}\
  \bibnamefont {Mo}}, \bibinfo {author} {\bibfnamefont {Z.}~\bibnamefont
  {Hussain}}, \bibinfo {author} {\bibfnamefont {Z.-X.}\ \bibnamefont {Shen}},
  \bibinfo {author} {\bibfnamefont {F.}~\bibnamefont {Wang}}, \bibinfo {author}
  {\bibfnamefont {S.~G.}\ \bibnamefont {Louie}}, \ and\ \bibinfo {author}
  {\bibfnamefont {M.~F.}\ \bibnamefont {Crommie}},\ }\href
  {http://dx.doi.org/10.1038/nmat4061} {\bibfield  {journal} {\bibinfo
  {journal} {Nat Mater}\ }\textbf {\bibinfo {volume} {13}},\ \bibinfo {pages}
  {1091} (\bibinfo {year} {2014})}\BibitemShut {NoStop}%
\bibitem [{\citenamefont {Zhu}\ \emph {et~al.}(2015)\citenamefont {Zhu},
  \citenamefont {Chen},\ and\ \citenamefont {Cui}}]{Zhu15}%
  \BibitemOpen
  \bibfield  {author} {\bibinfo {author} {\bibfnamefont {B.}~\bibnamefont
  {Zhu}}, \bibinfo {author} {\bibfnamefont {X.}~\bibnamefont {Chen}}, \ and\
  \bibinfo {author} {\bibfnamefont {X.}~\bibnamefont {Cui}},\ }\href
  {http://dx.doi.org/10.1038/srep09218} {\bibfield  {journal} {\bibinfo
  {journal} {Sci. Rep.}\ }\textbf {\bibinfo {volume} {5}} (\bibinfo {year}
  {2015})}\BibitemShut {NoStop}%
\bibitem [{\citenamefont {Wu}\ \emph {et~al.}(2015)\citenamefont {Wu},
  \citenamefont {Qu},\ and\ \citenamefont {MacDonald}}]{Wu15}%
  \BibitemOpen
  \bibfield  {author} {\bibinfo {author} {\bibfnamefont {F.}~\bibnamefont
  {Wu}}, \bibinfo {author} {\bibfnamefont {F.}~\bibnamefont {Qu}}, \ and\
  \bibinfo {author} {\bibfnamefont {A.~H.}\ \bibnamefont {MacDonald}},\ }\href
  {\doibase 10.1103/PhysRevB.91.075310} {\bibfield  {journal} {\bibinfo
  {journal} {Phys. Rev. B}\ }\textbf {\bibinfo {volume} {91}},\ \bibinfo
  {pages} {075310} (\bibinfo {year} {2015})}\BibitemShut {NoStop}%
\bibitem [{\citenamefont {Berkelbach}\ \emph {et~al.}(2013)\citenamefont
  {Berkelbach}, \citenamefont {Hybertsen},\ and\ \citenamefont
  {Reichman}}]{Berkelbach13}%
  \BibitemOpen
  \bibfield  {author} {\bibinfo {author} {\bibfnamefont {T.~C.}\ \bibnamefont
  {Berkelbach}}, \bibinfo {author} {\bibfnamefont {M.~S.}\ \bibnamefont
  {Hybertsen}}, \ and\ \bibinfo {author} {\bibfnamefont {D.~R.}\ \bibnamefont
  {Reichman}},\ }\href {\doibase 10.1103/PhysRevB.88.045318} {\bibfield
  {journal} {\bibinfo  {journal} {Phys. Rev. B}\ }\textbf {\bibinfo {volume}
  {88}},\ \bibinfo {pages} {045318} (\bibinfo {year} {2013})}\BibitemShut
  {NoStop}%
\bibitem [{\citenamefont {Xiao}\ \emph {et~al.}(2010)\citenamefont {Xiao},
  \citenamefont {Chang},\ and\ \citenamefont {Niu}}]{Xiao10}%
  \BibitemOpen
  \bibfield  {author} {\bibinfo {author} {\bibfnamefont {D.}~\bibnamefont
  {Xiao}}, \bibinfo {author} {\bibfnamefont {M.-C.}\ \bibnamefont {Chang}}, \
  and\ \bibinfo {author} {\bibfnamefont {Q.}~\bibnamefont {Niu}},\ }\href
  {\doibase 10.1103/RevModPhys.82.1959} {\bibfield  {journal} {\bibinfo
  {journal} {Rev. Mod. Phys.}\ }\textbf {\bibinfo {volume} {82}},\ \bibinfo
  {pages} {1959} (\bibinfo {year} {2010})}\BibitemShut {NoStop}%
\bibitem [{\citenamefont {Neupert}\ \emph {et~al.}(2013)\citenamefont
  {Neupert}, \citenamefont {Chamon},\ and\ \citenamefont {Mudry}}]{Neupert13}%
  \BibitemOpen
  \bibfield  {author} {\bibinfo {author} {\bibfnamefont {T.}~\bibnamefont
  {Neupert}}, \bibinfo {author} {\bibfnamefont {C.}~\bibnamefont {Chamon}}, \
  and\ \bibinfo {author} {\bibfnamefont {C.}~\bibnamefont {Mudry}},\ }\href
  {\doibase 10.1103/PhysRevB.87.245103} {\bibfield  {journal} {\bibinfo
  {journal} {Phys. Rev. B}\ }\textbf {\bibinfo {volume} {87}},\ \bibinfo
  {pages} {245103} (\bibinfo {year} {2013})}\BibitemShut {NoStop}%
\bibitem [{\citenamefont {Provost}\ and\ \citenamefont
  {Vallee}(1980)}]{Provost80}%
  \BibitemOpen
  \bibfield  {author} {\bibinfo {author} {\bibfnamefont {J.~P.}\ \bibnamefont
  {Provost}}\ and\ \bibinfo {author} {\bibfnamefont {G.}~\bibnamefont
  {Vallee}},\ }\href {http://projecteuclid.org/euclid.cmp/1103908308}
  {\bibfield  {journal} {\bibinfo  {journal} {Comm. Math. Phys.}\ }\textbf
  {\bibinfo {volume} {76}},\ \bibinfo {pages} {289} (\bibinfo {year}
  {1980})}\BibitemShut {NoStop}%
\bibitem [{\citenamefont {Xiao}\ \emph {et~al.}(2007)\citenamefont {Xiao},
  \citenamefont {Yao},\ and\ \citenamefont {Niu}}]{Xiao07}%
  \BibitemOpen
  \bibfield  {author} {\bibinfo {author} {\bibfnamefont {D.}~\bibnamefont
  {Xiao}}, \bibinfo {author} {\bibfnamefont {W.}~\bibnamefont {Yao}}, \ and\
  \bibinfo {author} {\bibfnamefont {Q.}~\bibnamefont {Niu}},\ }\href {\doibase
  10.1103/PhysRevLett.99.236809} {\bibfield  {journal} {\bibinfo  {journal}
  {Phys. Rev. Lett.}\ }\textbf {\bibinfo {volume} {99}},\ \bibinfo {pages}
  {236809} (\bibinfo {year} {2007})}\BibitemShut {NoStop}%
\bibitem [{\citenamefont {Yao}\ \emph {et~al.}(2008)\citenamefont {Yao},
  \citenamefont {Xiao},\ and\ \citenamefont {Niu}}]{Yao08}%
  \BibitemOpen
  \bibfield  {author} {\bibinfo {author} {\bibfnamefont {W.}~\bibnamefont
  {Yao}}, \bibinfo {author} {\bibfnamefont {D.}~\bibnamefont {Xiao}}, \ and\
  \bibinfo {author} {\bibfnamefont {Q.}~\bibnamefont {Niu}},\ }\href {\doibase
  10.1103/PhysRevB.77.235406} {\bibfield  {journal} {\bibinfo  {journal} {Phys.
  Rev. B}\ }\textbf {\bibinfo {volume} {77}},\ \bibinfo {pages} {235406}
  (\bibinfo {year} {2008})}\BibitemShut {NoStop}%
\bibitem [{\citenamefont {Garate}\ and\ \citenamefont
  {Franz}(2011)}]{Garate11}%
  \BibitemOpen
  \bibfield  {author} {\bibinfo {author} {\bibfnamefont {I.}~\bibnamefont
  {Garate}}\ and\ \bibinfo {author} {\bibfnamefont {M.}~\bibnamefont {Franz}},\
  }\href {\doibase 10.1103/PhysRevB.84.045403} {\bibfield  {journal} {\bibinfo
  {journal} {Phys. Rev. B}\ }\textbf {\bibinfo {volume} {84}},\ \bibinfo
  {pages} {045403} (\bibinfo {year} {2011})}\BibitemShut {NoStop}%
\bibitem [{\citenamefont {Welton}(1948)}]{Welton48}%
  \BibitemOpen
  \bibfield  {author} {\bibinfo {author} {\bibfnamefont {T.~A.}\ \bibnamefont
  {Welton}},\ }\href {\doibase 10.1103/PhysRev.74.1157} {\bibfield  {journal}
  {\bibinfo  {journal} {Phys. Rev.}\ }\textbf {\bibinfo {volume} {74}},\
  \bibinfo {pages} {1157} (\bibinfo {year} {1948})}\BibitemShut {NoStop}%
\bibitem [{\citenamefont {Marzari}\ and\ \citenamefont
  {Vanderbilt}(1997)}]{Marzari97}%
  \BibitemOpen
  \bibfield  {author} {\bibinfo {author} {\bibfnamefont {N.}~\bibnamefont
  {Marzari}}\ and\ \bibinfo {author} {\bibfnamefont {D.}~\bibnamefont
  {Vanderbilt}},\ }\href {\doibase 10.1103/PhysRevB.56.12847} {\bibfield
  {journal} {\bibinfo  {journal} {Phys. Rev. B}\ }\textbf {\bibinfo {volume}
  {56}},\ \bibinfo {pages} {12847} (\bibinfo {year} {1997})}\BibitemShut
  {NoStop}%
\bibitem [{\citenamefont {Ma}\ \emph {et~al.}(2010)\citenamefont {Ma},
  \citenamefont {Chen}, \citenamefont {Fan},\ and\ \citenamefont {Liu}}]{Ma10}%
  \BibitemOpen
  \bibfield  {author} {\bibinfo {author} {\bibfnamefont {Y.-Q.}\ \bibnamefont
  {Ma}}, \bibinfo {author} {\bibfnamefont {S.}~\bibnamefont {Chen}}, \bibinfo
  {author} {\bibfnamefont {H.}~\bibnamefont {Fan}}, \ and\ \bibinfo {author}
  {\bibfnamefont {W.-M.}\ \bibnamefont {Liu}},\ }\href {\doibase
  10.1103/PhysRevB.81.245129} {\bibfield  {journal} {\bibinfo  {journal} {Phys.
  Rev. B}\ }\textbf {\bibinfo {volume} {81}},\ \bibinfo {pages} {245129}
  (\bibinfo {year} {2010})}\BibitemShut {NoStop}%
\bibitem [{\citenamefont {Zanardi}\ \emph {et~al.}(2007)\citenamefont
  {Zanardi}, \citenamefont {Giorda},\ and\ \citenamefont
  {Cozzini}}]{Zanardi07}%
  \BibitemOpen
  \bibfield  {author} {\bibinfo {author} {\bibfnamefont {P.}~\bibnamefont
  {Zanardi}}, \bibinfo {author} {\bibfnamefont {P.}~\bibnamefont {Giorda}}, \
  and\ \bibinfo {author} {\bibfnamefont {M.}~\bibnamefont {Cozzini}},\ }\href
  {\doibase 10.1103/PhysRevLett.99.100603} {\bibfield  {journal} {\bibinfo
  {journal} {Phys. Rev. Lett.}\ }\textbf {\bibinfo {volume} {99}},\ \bibinfo
  {pages} {100603} (\bibinfo {year} {2007})}\BibitemShut {NoStop}%
\bibitem [{\citenamefont {Liu}\ \emph {et~al.}(2013)\citenamefont {Liu},
  \citenamefont {Shan}, \citenamefont {Yao}, \citenamefont {Yao},\ and\
  \citenamefont {Xiao}}]{Liu13}%
  \BibitemOpen
  \bibfield  {author} {\bibinfo {author} {\bibfnamefont {G.-B.}\ \bibnamefont
  {Liu}}, \bibinfo {author} {\bibfnamefont {W.-Y.}\ \bibnamefont {Shan}},
  \bibinfo {author} {\bibfnamefont {Y.}~\bibnamefont {Yao}}, \bibinfo {author}
  {\bibfnamefont {W.}~\bibnamefont {Yao}}, \ and\ \bibinfo {author}
  {\bibfnamefont {D.}~\bibnamefont {Xiao}},\ }\href {\doibase
  10.1103/PhysRevB.88.085433} {\bibfield  {journal} {\bibinfo  {journal} {Phys.
  Rev. B}\ }\textbf {\bibinfo {volume} {88}},\ \bibinfo {pages} {085433}
  (\bibinfo {year} {2013})}\BibitemShut {NoStop}%
\bibitem [{\citenamefont {Qiu}\ \emph {et~al.}(2013)\citenamefont {Qiu},
  \citenamefont {da~Jornada},\ and\ \citenamefont {Louie}}]{Qiu13}%
  \BibitemOpen
  \bibfield  {author} {\bibinfo {author} {\bibfnamefont {D.~Y.}\ \bibnamefont
  {Qiu}}, \bibinfo {author} {\bibfnamefont {F.~H.}\ \bibnamefont {da~Jornada}},
  \ and\ \bibinfo {author} {\bibfnamefont {S.~G.}\ \bibnamefont {Louie}},\
  }\href {\doibase 10.1103/PhysRevLett.111.216805} {\bibfield  {journal}
  {\bibinfo  {journal} {Phys. Rev. Lett.}\ }\textbf {\bibinfo {volume} {111}},\
  \bibinfo {pages} {216805} (\bibinfo {year} {2013})}\BibitemShut {NoStop}%
\bibitem [{\citenamefont {Yu}\ \emph {et~al.}(2014)\citenamefont {Yu},
  \citenamefont {Liu}, \citenamefont {Gong}, \citenamefont {Xu},\ and\
  \citenamefont {Yao}}]{Yu14}%
  \BibitemOpen
  \bibfield  {author} {\bibinfo {author} {\bibfnamefont {H.}~\bibnamefont
  {Yu}}, \bibinfo {author} {\bibfnamefont {G.-B.}\ \bibnamefont {Liu}},
  \bibinfo {author} {\bibfnamefont {P.}~\bibnamefont {Gong}}, \bibinfo {author}
  {\bibfnamefont {X.}~\bibnamefont {Xu}}, \ and\ \bibinfo {author}
  {\bibfnamefont {W.}~\bibnamefont {Yao}},\ }\href
  {http://dx.doi.org/10.1038/ncomms4876} {\bibfield  {journal} {\bibinfo
  {journal} {Nat Commun}\ }\textbf {\bibinfo {volume} {5}} (\bibinfo {year}
  {2014})}\BibitemShut {NoStop}%
\end{thebibliography}
%%%%%%%

\section{Appendix A.}

Consider a ($N$-1)-point discretization of a smooth curve $\mathcal{C}$ in $\mathbf{k}$-space parametrized by points $\mathbf{k}_1, \mathbf{k}_2, \ldots, \mathbf{k}_{N-1}$. We want to show that Eq. (3) in main text holds up to second order in $d\mathbf{k}$. As we wish to express the Bloch overlap, $s_{\mathcal{C}}$ as an integral over $\mathcal{C}$, we take the logarithm of LHS to get,
\begin{equation}
\mathrm{ln} (s_\mathcal{C}) = \sum_{\mathcal{C}} \mathrm{ln} \langle u_{\mathbf{k}_{i}}| u_{\mathbf{k}_{i+1}} \rangle.
\end{equation}
In the continuum limit,
\begin{equation}
\mathrm{ln} \langle u_{\mathbf{k}}|u_{\mathbf{k}+d\mathbf{k}} \rangle = \mathrm{ln} (1 + \langle u|\partial_{i} | u \rangle dk_i + \frac{1}{2} \langle u | \partial_{i} \partial_{j} | u \rangle dk_i dk_j + \ldots),
\end{equation}
where partial derivatives are taken in $\mathbf{k}$-space and summation over indices is assumed. Using $\mathrm{ln} (1+x)$ = $x - \frac{x^2}{2} + \ldots$ with $x$ = $\langle u|\partial_{i} | u \rangle dk_i + \frac{1}{2} \langle u | \partial_{i} \partial_{j} | u \rangle dk_i dk_j$, we get up to second order,
\begin{align}
\mathrm{ln} \langle u_{\mathbf{k}}|u_{\mathbf{k}+d\mathbf{k}} \rangle = \langle u|\partial_{i} | u \rangle dk_i + \frac{1}{2} \langle u | \partial_{i} \partial_{j} | u \rangle dk_i dk_j \\
- \frac{1}{2} \langle u|\partial_{i} | u \rangle \langle u|\partial_{j} | u \rangle dk_i dk_j + \ldots), \nonumber
\end{align}
With the above identity, $s_c$ becomes,
\begin{eqnarray}
s_{\mathcal{C}} \sim &\mathrm{exp}& \left[\oint_{\mathcal{C}} \langle u | \partial_i | u \rangle dk_i \right] \\
&\times& \mathrm{exp}\left[\frac{1}{2}\oint_{\mathcal{C}}(\langle u | \partial_{i} \partial_{j} | u \rangle - \langle u|\partial_{i} | u \rangle \langle u|\partial_{j} | u \rangle) dk_i dk_j \right] \ldots \nonumber
\end{eqnarray}
Using $\mathcal{A}_i$ = $-i \langle u | \partial_i | u \rangle$ and $\oint_\mathcal{C} \mathcal{A}_i dk_i$ = $\int \Omega dk_i \wedge dk_j$ we get,
\begin{equation}
s_{\mathcal{C}} \sim \mathrm{exp}\left[ i \int \Omega dk_i \wedge dk_j \right] \mathrm{exp}\left[\frac{1}{2}\oint_{\mathcal{C}} \left( \langle u | \partial_i \partial_j | u \rangle + \mathcal{A}_i \mathcal{A}_j\right) dk_i dk_j \right]\ldots
\end{equation}
Under local $U(1)$ transformation $\tilde{u}_{\mathbf{k}}$ = $u_{\mathbf{k}} e^{i \alpha (\mathbf{k})}$,
\begin{equation}
\langle \tilde{u} | \partial_i \partial_j | \tilde{u} \rangle + \tilde{\mathcal{A}}_i \tilde{\mathcal{A}}_j \rightarrow \langle u | \partial_i \partial_j | u \rangle + \mathcal{A}_i \mathcal{A}_j + i \partial_i \partial_j \alpha.
\end{equation}
As $s_{\mathcal{C}}$ is gauge-invariant under the transformation, when integral over closed $\mathcal{C}$ is preformed, the gauge-dependent term $i\partial_i \partial_j \alpha$ must vanish. Evidently, the real part of the above equation is gauge-invariant as $\mathcal{A}_i$s are real. Therefore, the following should hold,
\begin{eqnarray}
s_{\mathcal{C}} = &\mathrm{exp}&\left[i \int \mathbf{\Omega} \cdot d\mathbf{S}_{\mathcal{C}}\right] \\
&\times& \mathrm{exp}\left[\frac{1}{2}\oint_{\mathcal{C}}\left(\mathrm{Re}\left[ \langle u | \partial_i \partial_j | u \rangle\right] + \mathcal{A}_i \mathcal{A}_j \right)dk_i dk_j \right] \ldots \nonumber
\end{eqnarray}
Using $\mathrm{Re}\left[ \langle u | \partial_i \partial_j | u \rangle\right]$ = $-\mathrm{Re}\left[ \langle  \partial_i u |\partial_j  u \rangle\right]$, we arrive at Eq. (3) of main text.

\end{document}